\documentclass[prl,twocolumn,aps,byrevtex,showpacs,tightenlines]{revtex4}

\begin{document}
\title{The Partition Ensemble Fallacy Fallacy}
\author{Kae Nemoto, and Samuel L. Braunstein}
\affiliation{Informatics, Bangor University, Bangor LL57 1UT, UK}
\date{\today}

\begin{abstract}
The Partition Ensemble Fallacy was recently applied to claim no
quantum coherence exists in coherent states produced by lasers.
We show that this claim relies on an untestable belief of a particular 
prior distribution of absolute phase. One's choice for the prior
distribution for an unobservable quantity is a matter of `religion'. 
We call this principle the {\it Partition Ensemble Fallacy
Fallacy\/}. Further, we show an alternative approach to construct 
a relative-quantity Hilbert subspace where unobservability of certain
quantities is guaranteed by global conservation laws. This approach is
applied to coherent states and constructs an approximate
relative-phase Hilbert subspace.
\end{abstract}

\pacs{03.67.Lx, 42.50.Dv, 89.70.+c}
\maketitle

\section{Introduction}
The representation of a state and its associated interpretation 
are fundamental issues in quantum mechanics. For example, there are
typically an infinite number of ensembles $\{P_j\}$ with which one
may decompose a general state $\rho$ via
\begin{equation}
\rho = \sum_j p_j P_j\;, \quad p_j\ge 0\;,\quad P_j^2 = P_j \;.
\end{equation}
Only when the state is pure, does this representation become unique.
The laws of quantum mechanics say that (in the absence of any additional
information other than the state's identity) no physical 
interpretation can be based on a preferred choice of an ensemble
for this decomposition \cite{Kok00}. This result has been coined
the {\it Partition Ensemble Fallacy\/} (PEF) \cite{Kok00}.

Recently, this principle has been used to attack the meaning of
coherent states in quantum mechanics \cite{Rudolph01}. It has long been 
argued, though without rigorous proof, that the absolute phase of an 
electromagnetic field is not observable \cite{Molmer97,Gea90}. Thus,
it has been asserted that the nominal description of light from a laser 
as being a coherent state $\big|\,|\alpha|e^{-i\phi}\big\rangle$ should 
be averaged over the unknowable quantity $\phi$ \cite{Rudolph01}. 
The resulting description of the coherent state then becomes
\begin{eqnarray} \label{phi}
\rho_{\strut \rm PEF}
&=& \int_0^{2\pi} \frac{d\phi}{2\pi} P(\phi) \big|\, |\alpha|
e^{-i\phi}\big\rangle\big\langle |\alpha| e^{-i\phi} \big| \\
\label{fock}
&=& e^{-|\alpha|^2}
\sum_{n=0}^{\infty}\frac{|\alpha|^{2n}}{n!}|n\rangle\langle n| \;,
\end{eqnarray}
where following Ref.~\cite{Rudolph01} we have taken the prior probability
$P(\phi)$ to be flat. Thus, the ensemble of states being produced by
a laser could as easily be chosen as number states instead of coherent 
states with fixed coherent amplitude $|\alpha|$. Recalling that the PEF 
disallows interpretations for states based on a preferred choice,
we should infer that experiments using lasers cannot be reliably interpreted as 
demonstrating features or properties of coherent states. This is the
logic behind the argument of Rudolph and Sanders \cite{Rudolph01}.

The automatic assumption that the prior distribution of phases $P(\phi)$ 
should be taken as flat appears straightforward. Ordinarily, when one has 
an unknown quantity, one assigns a prior distribution based on whatever 
prior information is available. If one lacks {\it any\/} information then 
one tries to rely on symmetries in the problem. Thus, since any choice
of absolute phase $\phi$ leads to the same observable results the flat
prior appears to be the canonical choice. However, there is something
fishy about this reasoning. That a prior distribution is a meaningful 
summary of our knowledge (or lack thereof) depends on the full procedure
of inference (for example, Bayesian). Here we are talking about a 
quantity which is not simply unknown, but unknowable. No inference can
ever be made about it based on new information. In fact, 
{\it absolutely any\/} choice can be made for $P(\phi)$ and that choice
is completely untestable. All predictions one could make would agree
whatever choice one had for $P(\phi)$. To this extent, one's choice for
the prior distribution for an unobservable quantity is a matter of `religion.'
It lies outside the realm of science.

By contrast to the choice for a prior made by Rudolph and Sanders, a
choice consisting of a delta function would make the entire
application of the PEF inadmissible (since we would be dealing with
pure states). However, since the application of a principle cannot
depend upon an arbitrary (and untestable) choice it clearly follows
that the PEF does not apply here. We call this principle the {\it
Partition Ensemble Fallacy Fallacy\/} (PEFF).

If the absolute phase is truly unobservable then the PEFF guarantees
experimentalists the freedom (of `religion') to continue talking about
coherent states as a state of the form of (\ref{phi}). 
Even if one prepares a known state (\ref{fock}) as if to be the laser
output, a coherent state is still applicable. The PEFF simply says
that the usual coherent state language is unfalsifiable. 
Mathematically, the freedom to choose the prior due to the
unobservability of $\phi$ induces an equivalent relation between all
states (\ref{phi}).  This indicates that this freedom of `language'
imposes the absence of the physical meaning of purity (or mixture) on
the state representation.  
This suggests that a single mode state fails to represent the
quantum nature of the system due to the unobservability of absolute
phase.

Since this argument relies on the unproved unobservability of $\phi$, 
we may consider the validity of the coherent-state language in
the absence of the unobservability of absolute phase. As we have seen,
as long as we deal with a single mode state, the free choice of
`language' always exists.  This implies that to construct a state
which is free from the prior, the single mode representation has to be
abandoned.  As result, the difference
between the unobservability of absolute phase and the untestable prior
would not make a big difference in the requirement for state
representation in quantum mechanics. 
The state representation for the laser output might have to involve 
multimode to serve the quantum nature of the system.  In this paper we
investigate this in detail.

We start by considering much simpler situations where variables are
guaranteed to be unobservable due to global symmetries and the
Wigner-Araki-Yanase (WAY) theorem. We shall see that reasoning of
Rudolph and Sanders would leave most of the formalism of quantum mechanics
as unavailable if generally applied to unobservable quantities. Typically,
in these cases, an alternative approach based on relative variables
allows one to circumvent the entire discussion about unobserved quantities.
It appears impossible to construct an exact relative-phase Hilbert
subspace, however we give an explicit construction of an approximate
relative-phase Hilbert space. The advantage of this formulation,
despite the added complication, is that it allows the usual coherent
state language to be used without assuming that phase is unobservable.

\section{the Wigner-Araki-Yanase theorem}
The Wigner-Araki-Yanase (WAY) theorem gives us a playground of
systems where unobservability of certain quantities is guaranteed by
global conservation laws. The WAY theorem states that any operator
which does not commute with an operator of the global conservation is
not observable \cite{Wigner52, Araki60}. Consider a system consists of
the observed subsystem and its measuring apparatus. As the total
momentum $\hat{\Pi}$ of the system is conserved, a position operator
$\hat{x}$ of the observed subsystem is unobservable. This is because
the position operator does not commute with the total momentum and
such measurement process violates the conservation law. Application of
Rudolph and Sanders reasoning to this example results in an arbitrary
position eigenstate to be   
\begin{equation}\label{intp}
\int dX P(X) \hat{D}(X)|x\rangle\langle x|\hat{D}^{\dagger}(X)
= \int dp |p\rangle \langle p| = {\hat 1} \label{unity},
\end{equation} 
where $\hat{D}=e^{-iX\hat{\Pi}}$ is the position displacement operator.

As the position $\hat{x}$ is unobservable and hence $P(X)$ is
completely arbitrary, all states of the left hand
side of (\ref{intp}) are equivalent. The freedom to choose the prior
is to be shown by expectation values of all possible
observables. The total momentum conservation restricts Hamiltonian and
time evolution unitary operator to commute with the total
momentum, that is $\hat{U}\hat{\Pi}\hat{U}^{\dagger}=\hat{\Pi}$.  The
expectation of an arbitrary observable $\langle{\hat A}\rangle$ is
given by 
\begin{eqnarray}\label{expectation}
\langle\hat{A}\rangle &=&
\mbox{Tr}[\hat{A}\hat{U}^{\dagger}\rho\hat{U}]\nonumber\\
&=& \int P(X) dX \langle x| \hat{U}\hat{A}\hat{U}^{\dagger}
|x\rangle.\\
&=& \langle \phi  \nonumber
| \hat{U}\hat{A}\hat{U}^{\dagger} | \phi\rangle.
\end{eqnarray}
The prior probability distribution $P(X)$ is completely arbitrary for any
physical quantities associated with the system. Therefore we can
use the pure-state language to consistently treat the system.

\section{relative quantities and state preparation}
Now we give a way to construct relative-quantity Hilbert space.
According to the WAY theorem, a relative quantity of the unobservable
absolute operators can be observable. For example, a relative
position operator of the observed system to the apparatus
$\hat{x}_1-\hat{x}_2$ ($=\hat{x}_r$) commutes with the total momentum
and hence is observable, where $\hat{x}_1$ and $\hat{x}_2$ are the
absolute positions of the observed system and the apparatus respectively. 
We take eigenstates of an operator $\hat{x}_a=\hat{x}_1+\hat{x}_2$ to
construct the entire Hilbert space together with eigenstates of
$\hat{x}_r$. The Hilbert space for the entire system (the observed
system and the apparatus) can be expanded by
$\{|x_r\rangle\otimes|x_a\rangle\}$ as well as by 
$\{|x_1\rangle\otimes|x_2\rangle\}$. To construct a relative-position
Hilbert space, we start with separable states given by 
\begin{eqnarray} \label{separable}
\begin{array}{ll}
|\psi\rangle = |\psi_r\rangle \otimes |\psi_a\rangle, \mbox{ where} & 
\left\{ \begin{array}{l}
		 |\psi_r\rangle=\int dx_r  \psi_r(x_r)|x_r\rangle\\
		|\psi_a\rangle=\int dx_a \psi_a(x_a)|x_a\rangle.\!
	\end{array}
\right.
\end{array}\!\!\!
\end{eqnarray}
Here the state is separable in terms of the two subspaces of
$\{|x_r\rangle\}$ and $\{|x_a\rangle\}$.

Similarly to the procedure of Eq.\ (\ref{intp}), integrating the state
over the amount of displacement $X$ by the operator $\hat{D}(X)$ with
the prior distribution $P(X)$, we obtain 
\begin{equation} \label{XX}
\rho_{ra} = 
\int dX P(X) e^{-iX\hat{\Pi}} |\psi\rangle\langle \psi| e^{iX\hat{\Pi}}.
\end{equation}
However, the operator $\hat{x}_r$ commutes with the total momentum
$\Pi$, then the state $|\psi_r\rangle$ is preserved under the
action of the displacement operator. This allows the density matrix to be 
\begin{eqnarray} \label{sep}
\rho = |\psi_r\rangle\langle \psi_r| \otimes \rho_a.
\end{eqnarray}
where
\begin{eqnarray}
|\psi_r\rangle &=& \int dx_r \psi_r(x_r)|x_r\rangle\nonumber\\
\rho_a &=& \int\int\int dX P(X) dx_adx'_a
\psi_a(x_a)\psi^*_a(x'_a)\nonumber\\
&& \times e^{-iX\hat{\Pi}}|x_a\rangle\langle x'_a|
e^{iX\hat{\Pi}}.\nonumber
\end{eqnarray}
The state $|\psi_r\rangle$ is on the relative-position Hilbert space.
The relative-position Hilbert space is constructed to be completely
free from the prior distribution and the argument associated with the
unobservability.

By assuming arbitrary separable states (\ref{separable}), we naturally
include the case of non-separable states.  Some entangling operators
such as a SUM gate ($\exp \big(-i \hat{x}_r \otimes \hat{\Pi}\big)$)
commute with the total momentum and hence are allowed. These operators
with superpositions can generate entanglement. In this case, the state
has to be more generally represented by
\begin{equation}\label{entangled}
|\phi\rangle = \int dx_r dx_a \psi(x_r,x_a) |x_r, x_a\rangle.
\end{equation}
In the case where the state is entangled, the function $\psi(x_r,x_a)$
cannot be written as $\psi_r(x_r)\psi_a(x_a)$. If we take the same
procedure to this state, then we have 
\begin{eqnarray} \label{ent}
\rho &=& \int\cdots\int dx_a dx_{a'} dx_r dx_{r'}
\psi(x_r,x_a)\psi^*(x_r',x_a')|x_r\rangle \langle x_r'|\nonumber\\ 
&\otimes& \int dX P(X) |x_a+X\rangle \langle x_a'+X|.
\end{eqnarray}
It is unfortunately not trivial how a relative-position
Hilbert space can be extracted in this state. Next we will see a
consideration of state preparation under the conservation laws helps
us to construct a consistent relative-position Hilbert space.

As the total momentum is constant, any eigenstate of the total
momentum can be a state of the total-momentum Hilbert space.  A
superposition of the total momentum eigenstates is also consistent
with the constant total momentum requirement. As we have discussed
above, a superposition can generate entanglement with some entangling
operator, while an eigenstate of the total momentum cannot be
entangled with the relative-position subspace. With an eigenstate of
the total momentum, none of the operators which generate a
superposition of the eigenstates is allowed under the conservation
law. This leads to the necessity of a third system to be involved in
the state preparation process. When we consider the whole process of
measurement including state preparation, each eigenstate of the total
momentum is solely consistent with the conservation law. Furthermore,
it is inconsistent to treat the two processes, a measurement and a
state preparation, in different spaces. This means that even if the
system of the observe system and the apparatus recovers the
conservation of the total momentum after the state preparation, the 
system cannot completely eliminate the third system.  
A closed system with the momentum conservation is invariant in
transformation by its absolute position, so different values of the
total momentum gives the same state to the system.  Two different
values of the {\it total momentum} $\hat{\Pi}$ become distinct when these
are realized in the extended system.
Thus, the superposition should be considered to lie on a
relative-quantity subspace in the extended system.
For a closed system with the momentum conservation, as the eigenstate
of the total momentum is the only state consistent, any state can be
represented as (\ref{separable}) and hence the relative-position
subspace always can be constructed.

\section{Phase of laser light field}
Now we turn our attention back to the laser light field and apply our
procedure to this particular case. The expected unobservability of the
absolute phase is the motivation to introduce a relative phase of two
mode coherent state given by  
\begin{eqnarray} \label{ab}
|\alpha, \beta\rangle &=& \big|\; |\alpha| e^{-i\phi_{\alpha}}\rangle \otimes 
\big|\; |\beta|e^{-i\phi_{\beta}}\rangle \nonumber\\
&=& \sum_{n_1}^{\infty}\sum_{n_2}^{\infty}
\frac{\alpha ^{n_1}\beta^{n_2}}{\sqrt{n_1 ! n_2!}}|n_1,n_2\rangle.
\end{eqnarray}
The total photon number of the state is $N=n_1+n_2$ 
and the difference photon number is $M=(n_1-n_2)/2$
which is either integer (for even total photon numbers) or
half-integer (for odd total photon numbers).  The state (\ref{ab}) can
be alternatively expanded by the eigenstates characterized by these
quantum numbers $N$ and $M$ as
\begin{eqnarray} \label{twomode-nm}
|\alpha,\beta\rangle &=& e^{-\frac{|\alpha|^2+|\beta|^2}{2}} 
\sum_{N=0}^{\infty} \sum_{M=-N/2}^{N/2} \alpha^{\frac{N}{2}+M}
\beta^{\frac{N}{2}-M} \nonumber\\
&\times& \Big[\big(\frac{N}{2}+M\big)! \big(\frac{N}{2}-M\big)! 
\Big]^{-1/2} |N,M\rangle.
\end{eqnarray}
Obviously this state is not separable in terms of the two subspaces,
$\{|N\rangle\}$ and $\{|M\rangle\}$. In this case we cannot simply
extract the relative-phase subspace, so we require an ingredient to
approximately extract the relative-phase subspace. Taking a 
set of parameters as 
\begin{eqnarray}\label{para}
\left\{ \begin{array}{l}

\frac{|\alpha|}{\langle\hat{N}\rangle^{1/2}}=-\sin\frac{\theta}{2},\;
\frac{|\beta|}{\langle\hat{N}\rangle^{1/2}}=\cos\frac{\theta}{2}\\
	\phi_{\alpha}-\phi_{\beta}=\phi_r.\\
	\end{array}
\right.
\end{eqnarray}
The two mode coherent state can be written as the sum of spin coherent
states, yielding
\begin{eqnarray} \label{twomode}
|\alpha,\beta\rangle = e^{-\frac{\langle\hat{N}\rangle}{2}} 
\sum_{N=0}^{\infty}
\frac{\big(\langle\hat{N}\rangle^{1/2}e^{-i\phi_{\beta}}\big)^N}{\sqrt{N!}}
|N, \theta, \phi_r\rangle.
\end{eqnarray}
Here $|N, \theta, \phi_r\rangle$ is a spin-$N/2$ coherent state with
the parameterization (\ref{para}). 
Alternatively the spin coherent state may be parameterized by $\xi$ ($= -
\frac{|\alpha|}{|\beta|}e^{-i\phi_r}$) as
\begin{eqnarray}
|N,\xi\rangle =\!\!\! \sum_{M=-\frac{N}{2}}^{\frac{N}{2}}\!\!
\left( 
	\!\!\begin{array}{c}
	N\\
	\frac{N}{2}-M
	\end{array}\!\! \right)^{\!\frac{1}{2}}\!
(1+|\xi|^2)^{-\frac{N}{2}}\xi^{\frac{N}{2}+M} |N,M\rangle.
\end{eqnarray}

If we ignore the spin coherent space, then the state for the total
photon number can be realized as a coherent state of
$|\langle\hat{N}\rangle^{1/2} e^{-i\phi_{\beta}}\rangle$. 
When $\langle\hat{N}\rangle^{1/2}$ goes infinity, the contribution of
components for small $N$ to the sum is negligible and the main
contribution is the terms of the order $N \sim
\langle\hat{N}\rangle^{1/2}$. In the large limit of $N$, the spin
coherent state can be contracted to a Weyl-Heisenberg (WH) coherent state.
When $|\alpha|\sim |\beta|$, the state can be typically 
contracted to a WH coherent state, 
\begin{eqnarray}
|\theta,\phi_r\rangle \to |-\sqrt{2}|\alpha|e^{-i\phi_r}\rangle.
\end{eqnarray}
At the limit, this coherent state is approximately separable with the
subspace of the total photon number, hence an approximate
relative-phase subspace has been constructed. However this
approximation is not so useful as the limit brings $|\alpha|$ also to
infinity as $\sqrt{2}\alpha \sim \langle\hat{N}\rangle^{1/2}$.  
By contrast, when $|\alpha|<<|\beta|$ is satisfied, the
group contraction may be taken in the order of $\langle
\hat{N}\rangle$. In this case the spin state $|N, \theta,
\phi_r\rangle$ is contracted by a parameter $\epsilon = 1/|\beta|$
as
\begin{equation}
\xi=-\epsilon |\alpha|e^{-i\phi_r}, \; (\epsilon \to \infty).
\end{equation}
In this contraction, the spin size given by $|\beta|^2$ goes to
infinity with $\epsilon \to 0$ and the state is contracted to a
WH coherent state $|-|\alpha|e^{-i\phi_r}\rangle$.  

The coherent state from laser can be approximately represent as
\begin{equation} \label{relative}
|\alpha,\beta\rangle\langle\alpha,\beta| \sim |-|\alpha|e^{-i\phi_r}\rangle\langle -
|\alpha|e^{-i\phi_r}|\otimes \rho_N,
\end{equation}
under the condition 
\begin{equation}\label{cond}
\langle\hat{N}\rangle \sim |\beta|^2 >> |\alpha|^2.
\end{equation}
The coherent state is constructed in the subspace of the relative
phase.

To conclude, we have shown the explicit construction of an approximate
relative-phase Hilbert space. The two mode coherent state can be
represented as a pure coherent state in the relative-phase subspace
under the condition (\ref{cond}).  This state presentation of relative
phase is free from a choice of prior distribution, and hence
circumvents the entire discussion about unknowable absolute phase.

\vskip 0.1truein

SLB thanks Asher Peres and Howard Wiseman for discussions. 
This work is funded in part under project QUICOV as part of the
IST-FET-QJPC programme.

\end{document}